\documentclass[preprint]{aastex631}

\renewcommand{\thefootnote}{\textit{\alph{footnote}}}
\renewcommand{\thefootnote}{\fnsymbol{footnote}}

\renewcommand{\thefootnote}{\textit{\alph{footnote}}}
\usepackage{graphicx}
\usepackage{aas_macros,amsmath,amssymb,xcolor,url}
\usepackage{multirow}

\usepackage[nolist]{acronym}

\renewcommand{\thefootnote}{\textit{\alph{footnote}}}

\usepackage[symbol]{footmisc}
\renewcommand{\thefootnote}{\fnsymbol{footnote}}

\usepackage[none]{hyphenat}
\usepackage{multibib}
\newcites{New}{References}
\usepackage{cleveref}

\shorttitle{HW-accelerated Inference for Real-Time GW Astronomy}
\shortauthors{Gunny et al.}
\graphicspath{{./}{figures/}}
\begin{document}

\title{Hardware-accelerated Inference for Real-Time Gravitational-Wave Astronomy}

\author{Alec Gunny}
\affiliation{Department of Physics,\\
Massachusetts Institute of Technology,\\
Cambridge, Massachusetts 02139, USA}
\affiliation{LIGO Laboratory,\\
185 Albany St, MIT,\\
Cambridge, MA 02139, USA}
\affiliation{The NSF AI Institute 
for Artificial Intelligence
and Fundamental Interactions}

%\author[0000-0001-8411-9620]{Dylan Rankin}
\author{Dylan Rankin}
\affiliation{Department of Physics,\\
Massachusetts Institute of Technology,\\
Cambridge, Massachusetts 02139, USA}
\affiliation{The NSF AI Institute 
for Artificial Intelligence
and Fundamental Interactions}

\author{Jeffrey Krupa}
\affiliation{Department of Physics,\\
Massachusetts Institute of Technology,\\
Cambridge, Massachusetts 02139, USA}
\affiliation{The NSF AI Institute 
for Artificial Intelligence
and Fundamental Interactions}

\author{Muhammed Saleem}
\affiliation{School of Physics and Astronomy,\\
University of Minnesota,\\
Minneapolis, Minnesota 55455, USA}

\author{Tri Nguyen}
\affiliation{Department of Physics,\\
Massachusetts Institute of Technology,\\
Cambridge, Massachusetts 02139, USA}
\affiliation{LIGO Laboratory,\\
185 Albany St, MIT,\\
Cambridge, MA 02139, USA}
\affiliation{The NSF AI Institute 
for Artificial Intelligence
and Fundamental Interactions}

\author{Michael Coughlin}
\affiliation{School of Physics and Astronomy,\\
University of Minnesota,\\
Minneapolis, Minnesota 55455, USA}

\author{Philip Harris}
\affiliation{Department of Physics,\\
Massachusetts Institute of Technology,\\
Cambridge, Massachusetts 02139, USA}
\affiliation{The NSF AI Institute 
for Artificial Intelligence
and Fundamental Interactions}

\author{Erik Katsavounidis}
\affiliation{Department of Physics,\\
Massachusetts Institute of Technology,\\
Cambridge, Massachusetts 02139, USA}
\affiliation{LIGO Laboratory,\\
185 Albany St, MIT,\\
Cambridge, MA 02139, USA}

\author{Steven Timm}
\affiliation{Fermi National Accelerator Laboratory,\\
Batavia, IL 60510, USA}
\author{Burt Holzman}
\affiliation{Fermi National Accelerator Laboratory,\\
Batavia, IL 60510, USA}



\begin{abstract}
The field of transient astronomy has seen a revolution with the first gravitational-wave detections and the arrival of multi-messenger 
observations they enabled.
Transformed by the first detection of binary black hole and binary neutron star mergers, computational demands in gravitational-wave astronomy are expected to grow by at least a factor of two over the next five years as the global network of kilometer-scale interferometers are brought to design sensitivity.
With the increase in detector sensitivity, real-time delivery of gravitational-wave alerts will become increasingly important as an enabler of multi-messenger followup.
In this work, we report a novel implementation and deployment of deep learning inference for real-time gravitational-wave data denoising and astrophysical source identification.
This is accomplished using a generic Inference-as-a-Service model that is capable of adapting to the future needs of gravitational-wave data analysis.
Our implementation allows seamless incorporation of hardware accelerators and also enables the use of commercial or private (dedicated) as-a-service computing.
Based on our results, we propose a paradigm shift in low-latency and offline computing in gravitational-wave astronomy.
Such a shift can address key challenges in peak-usage, scalability and reliability, and provide a data analysis platform particularly optimized for deep learning applications.
The achieved sub-millisecond scale latency will also be relevant for any machine learning-based real-time control systems that may be invoked in the operation of near-future and next generation ground-based laser interferometers, as well as the front-end collection, distribution and processing of data from such instruments.
\end{abstract}


%\keywords{GPU computing (1969) --- Computational methods (1965)}


\section*{}
We have entered a new era where discoveries in astronomy will be driven by combining observations in gravitational waves, the electromagnetic spectrum, as well as neutrinos, e.g. Ref.~\citep{DiCo2020,AaAc2018}.
This is often referred to as multi-messenger astronomy (MMA) and it has been enabled by the direct detection of gravitational-wave transients~\citep{Abbott:2016blz,AbEA2017b} with the large ground-based laser interferometers LIGO~\citep{ligoHarry_2010,aLIGO,ligo2015} and Virgo~\citep{adVirgo}.  
The data analyses for gravitational-wave transients of both known and unknown signal morphology present a major computational challenge for these instruments as they prepare to enter their fourth observing run (referred to as ``O4'') in the summer of 2022~\citep{Abbott_2020_obs}.
Existing gravitational-wave detection algorithms for compact binary systems rely heavily on parameterized waveforms (templates) and the use of matched-filtering techniques, e.g. Refs~\citep{Usman:2015kfa,2020arXiv201005082C}.
Such approaches scale poorly with the expected low frequency improvement of instruments as well as with the expanding parameter space needed to cover spin effects and sub-solar mass compact binaries~\citep{Abbott_2019ss}.
The anticipated addition of KAGRA~\citep{kagraPhysRevD.88.043007,Som2012} in the international network of detectors observing the gravitational-wave sky is expected to further increase the computational demands.
%Deep-learning based detection algorithms are rising to meet this challenge~\citep{GeHu2018}.

At the same time, gravitational-wave transients of ill-defined morphology (e.g., supernovae, neutron star glitches and potentially yet-unknown astrophysical systems) are susceptible to instrumental and environmental noise that is hard to simulate and often challenging to subtract~\citep{ZeCo2017,DaAr2021}.
Such noise sources will affect searches for binary systems with sufficient signal duration that will make it statistically probable to overlap with non-Gaussian artifacts, such as in the case of GW170817~\citep{AbEA2017b}.
Detection of persistent gravitational-wave emission from known and unknown neutron stars or in the form of a stochastic radiation (although not a real-time detection process) may also be hindered by noise artifacts, mostly in the form of line noise appearing in the noise spectrum of the instruments~\citep{DaAr2021}.
While the vast majority of data analysis techniques employed in gravitational-wave searches are built on traditional time-frequency decomposition using the Fourier and other wavelet transforms, deep learning techniques have recently emerged as potentially powerful solutions to the computational challenges in this field.
Neural networks and other gradient-based learning algorithms have been proposed in gravitational-wave analyses for tasks like noise regression~\citep{VaHu2020,OrNg2020}, astrophysical searches~\citep{GeHu2018} and transient noise classification~\citep{ZeCo2017, MuAb2017}.
However, significant barriers exist before they can be used in large-scale gravitational-wave analyses, as large networks with complex architectures increase computational demands and incur large inference latencies.

For gravitational-wave astronomy, however, time is of the essence.
Astrophysical sources of transient gravitational waves, located at cosmological distances, are expected to be faint and fast-fading.
Thus, maximizing the time to search the sky for their counterparts is essential~\citep{Metzger:2016pju}.
Innovative hardware-based acceleration with Graphics Processing Units (GPUs) and Field Programmable Gate Arrays (FPGA) have recently been gaining ground within industry and academic research~\citep{10.1145/2788396,Duarte:2019fta} as methods for fast machine learning (ML) inference at large scales.
The computing landscape available for gravitational-wave searches mostly includes in-house computing clusters over which CPU-intensive workflows are run both for real-time as well as offline processing~\citep{CaNi2021}.
The rise of recent Heterogeneous High Performance Computing centers (HHPCs) has helped to illustrate an alternative computing model. 
The rapid growth of HHPCs provides a potential solution for a next-generation gravitational-wave computing model.

In order to take full advantage of accelerators, modifications must be made to the standard model of computing, in which pipelines directly manage the accelerated resources they use for execution.
An alternative model, which has gained popularity in other fields, is called ``as-a-service''~\citep{Krupa:2020bwg,Wang:2020fjr}. When used to specifically denote accelerated ML inference, it is referred to as Inference-as-a-Service (IaaS).
In this IaaS model, trained ML models are hosted in a centralized repository, from which an inference application loads and exposes them to requests from networked clients.
The user then requests inferences via a client by sending packets of inputs to the server via HTTP or gRPC (a remote process calling system often used in distributed computing).
The details of the server are abstracted from the user using standard client Application Programming Interfaces (APIs).
This allows the simple integration and management of heterogeneous computing resources as well as parallel and asynchronous execution of inference requests.

In the following, we will analyze the advantages of the IaaS paradigm using two deep learning models.
The first is DeepClean~\citep{OrNg2020}, a 1-D convolutional autoencoder able to predict and subtract noise present in the gravitational-wave sensing channel (referred to as ``strain''). 
Gravitational-wave interferometers are subject to technical and environmental noise that ultimately limit the instruments' ability to reach their design sensitivity. 
The bulk of data written onto disk by these instruments corresponds to auxiliary channels that monitor and record the interferometry as well as their physical environment.
This allows effective monitoring and regression of transient or continuous noise from the gravitational-wave measurement.
In order to achieve this, about 200,000 such auxiliary channels resulting in over 10\,MBps of time series data from each interferometer are continually written onto disk during normal data acquisition~\citep{AbAb2016}.
We generally refer to them as ``witness'' channels for their ability to record noise that may affect the measurement and their role in assisting noise subtraction from the strain channel.
DeepClean uses information from the auxiliary channels correlated with the strain channel in order to achieve noise reduction. 
It can also be customized to specific noise couplings for a variety of applications~\citep{OrNg2020}.
In the use-case described here, we use 21 auxiliary channels that are good witnesses of noise appearing in the strain channel and associated with monitor lines of power mains and their harmonics, including sidebands.

The second model we will use to demonstrate the IaaS framework, called BBHnet~\citep{tngetal-inprep}, is used in the archetypal search for compact binary black hole coalescences.
BBHnet utilizes 1-D convolutional neural networks in order to distinguish between binary black holes and detector noise, with the noise-regressed strain signal derived by DeepClean as its input.
The combination of DeepClean and BBHnet offers an end-to-end test of a pipeline that combines both data-cleaning and astrophysical searches.
The task of binary black hole identification is a challenging one, but deep learning has been shown to be an effective replacement~\citep{GeHu2018} for the template matching algorithms that are currently used.
The number of templates required for such algorithms can grow to the millions as searches expand to cover neutron stars and systems with sub-solar component masses~\citep{Abbott_2019ss}.
Moreover, the continuing improvement of the low-frequency sensitivity of interferometers and the addition of new detectors in the international gravitational-wave network~\citep{Abbott_2020_obs} also grow the template banks used in GW searches.
Deep learning is only expected to become more critical in this regime due to the large number of free parameters in the compact binary star system numerical problem.
At the same time, the ability to perform such gravitational-wave searches in real-time enables their follow-up via the electromagnetic spectrum and neutrinos, thus enabling multi-messenger astronomy.

To deploy these pipelines, we use an out-of-the-box inference service developed by NVIDIA called the Triton Inference Server~\citep{Triton}.
Triton supports concurrent inference execution on both CPUs and GPUs using multiple framework backends.
It is provided as a containerized application to make it portable to different deployment locations.
Triton automatically detects changes in the model repository from which it reads, and will update the models that it hosts in-memory in response to updates according to a prescribed versioning policy.
This ensures that all services, and their users, are kept in-sync with the latest developments and with one another. 
This is particularly beneficial for computing scenarios where a distributed user base is interested in accessing centrally-managed server resources.

\begin{figure}[htp]
    \centering
    \includegraphics[scale=0.7]{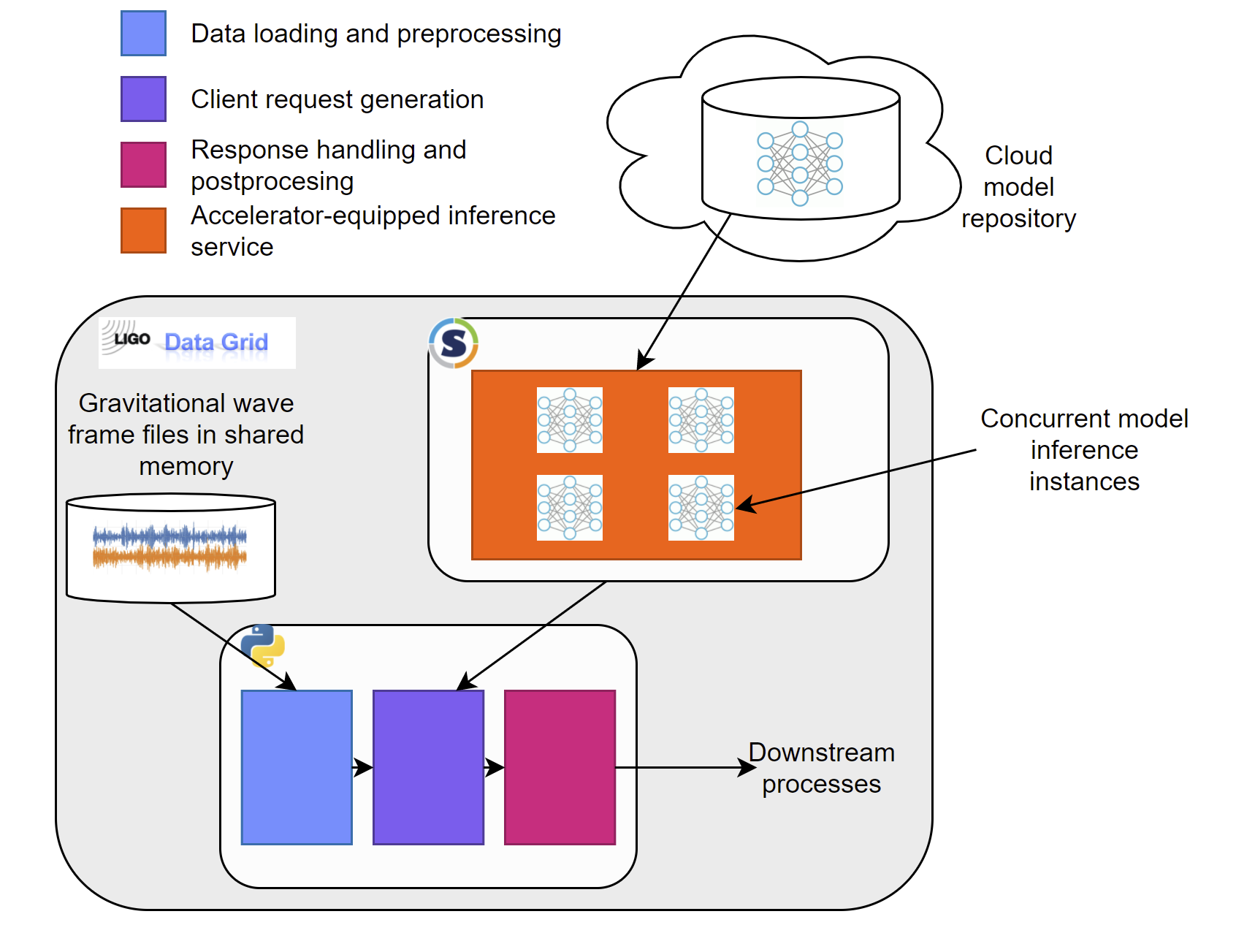}
    \caption{Example IaaS deployment scenario for the local online case. Locally (i.e., at the gravitational-wave detector sites) deployed client and server instances co-located with in-memory data sources to minimize latency. The server is deployed in a container using Singularity~\citep{singularity} and reads from a cloud-based model repository to stay in sync with updates.}
    \label{fig:triton-ldg}
\end{figure}

\begin{figure}[htb!]
\centering
    \includegraphics[scale=0.7]{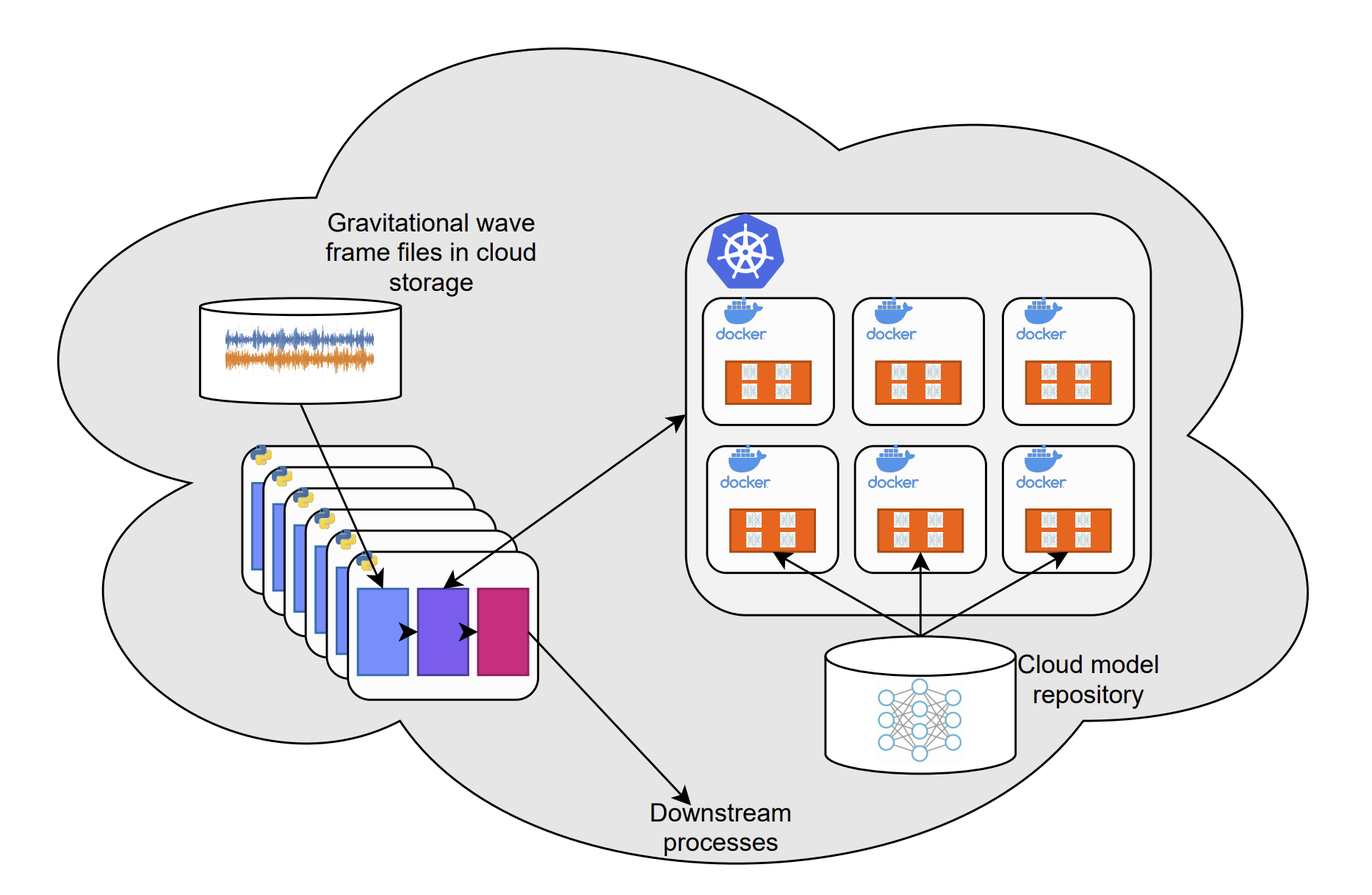}
    \caption{Example IaaS deployment scenario for the cloud-based offline case. Offline co-located cloud-based deployment where multiple server instances are managed by Kubernetes and data is split among multiple client virtual machines.}
    \label{fig:triton-cloud}
\end{figure}

Inference scenarios in gravitational-wave data analyses can be broadly separated into online and offline categories.
The online inference scenario requires low latency inferences for real-time processing of live streams during data collection runs.
The offline scenario, on the other hand, involves large-scale processing of archival data for use-cases such as model validation analyses or completion of transient event searches and corresponding catalogs following the definitive calibration of the instruments.
Fig.~\ref{fig:triton-ldg} and Fig.~\ref{fig:triton-cloud} depict the two IaaS deployment scenarios we have adopted for realistic online and offline use-cases in gravitational-wave experiments.
%TC:ignore

Fig.~\ref{fig:triton-ldg} depicts the deployment for an online pipeline which leverages DeepClean to remove noise from the strain channel at each detector to be made available to downstream transient event detection algorithms in real-time.
In order to meet the low-latency requirements of mutli-messenger astronomy, this pipeline is deployed fully on the LIGO Data Grid (LDG, \url{https://computing.docs.ligo.org/lscdatagridweb/}) so that the data sources, client, and inference service can minimize latency incurred by networked connections.
Fig.~\ref{fig:triton-cloud} presents a generic offline scenario we use for two archival data processing tasks.
These use-cases, unlike the online use-case, prioritize the total processing time for a given dataset instead of the individual request latency.
In addition, they can be massively parallelized by breaking the dataset into smaller, time-contiguous subsets which are assigned to individual client instances.
In both these scenarios, building an optimal IaaS pipeline involves tuning the same parameters, but their different objectives and constraints lead to drastically different decisions for which set of values is optimal.
Moreover, the streaming nature of gravitational-wave data presents unique challenges to the IaaS model in both deployment scenarios. In order to reduce the overall bandwidth going into the inference service, previous time series samples are cached on the GPU so that only the new samples need to be sent to the server. More details can be found in the methods section.

\begin{figure}[htbp!]
    \centering
    \includegraphics[scale=0.20]{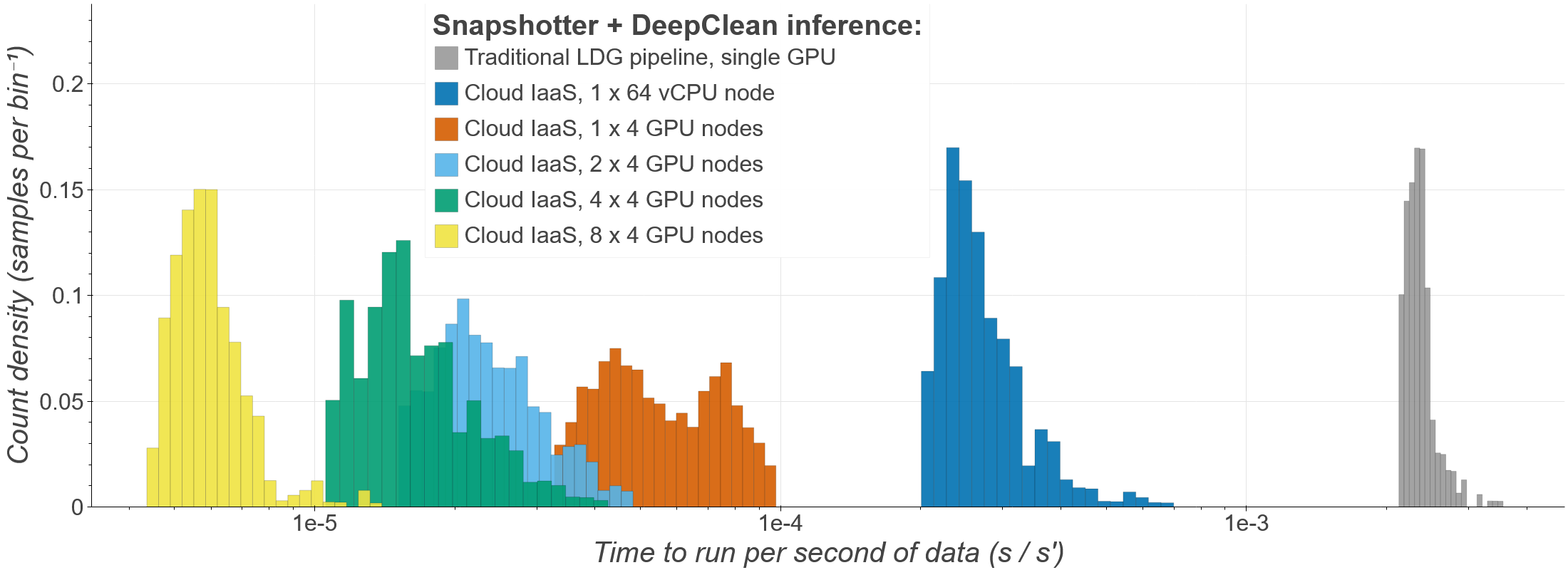}
    \caption{Distributions of the time required to process one second of data for DeepClean, $r = 0.25~\text{Hz}$. For these studies all GPU-equipped inference service nodes used NVIDIA V100 32GB GPUs and were equipped with 32 vCPUs and hosted 6 concurrent execution instances per GPU. The CPU-only inference service hosted 6 concurrent execution instances for the whole node.}
    \label{fig:dc_offline}
\end{figure}

As a first test of the offline use-case, we deploy DeepClean to remove noise from roughly a month's worth of strain data from the O3 observing run of the LIGO-Virgo instruments~\citep{Abbott_2020_obs}, using an inference sampling rate ($r$) of 0.25 Hz.
Figure~\ref{fig:dc_offline} shows the distribution of processing time per second of data achieved both by a traditional workflow on the LDG and by workflows leveraging an inference service on various amounts of cloud resources.
The LDG workflow, a traditional pipeline consisting of a single GPU, has the longest processing time, shown in grey.
For the IaaS workflow using only CPUs we observe a reduction in the processing time by a factor of 10 when compared to a traditional workflow, in spite of the additional sources of latency in the IaaS workflows.
This reduction comes largely from the inference service’s ability to execute multiple tasks concurrently, allowing for efficient parallelization of the work across multiple inference instances.
Further reductions in time are achieved by adding GPUs to the service.
An inference service equipped with 4 GPUs is able to decrease the processing time by another factor of 5, and the reduction continues proportionally as more service nodes are added.
These modifications to the number of GPUs can be handled seamlessly in the IaaS paradigm.

\begin{figure}[htb!]
\centering
\includegraphics[scale=0.7]{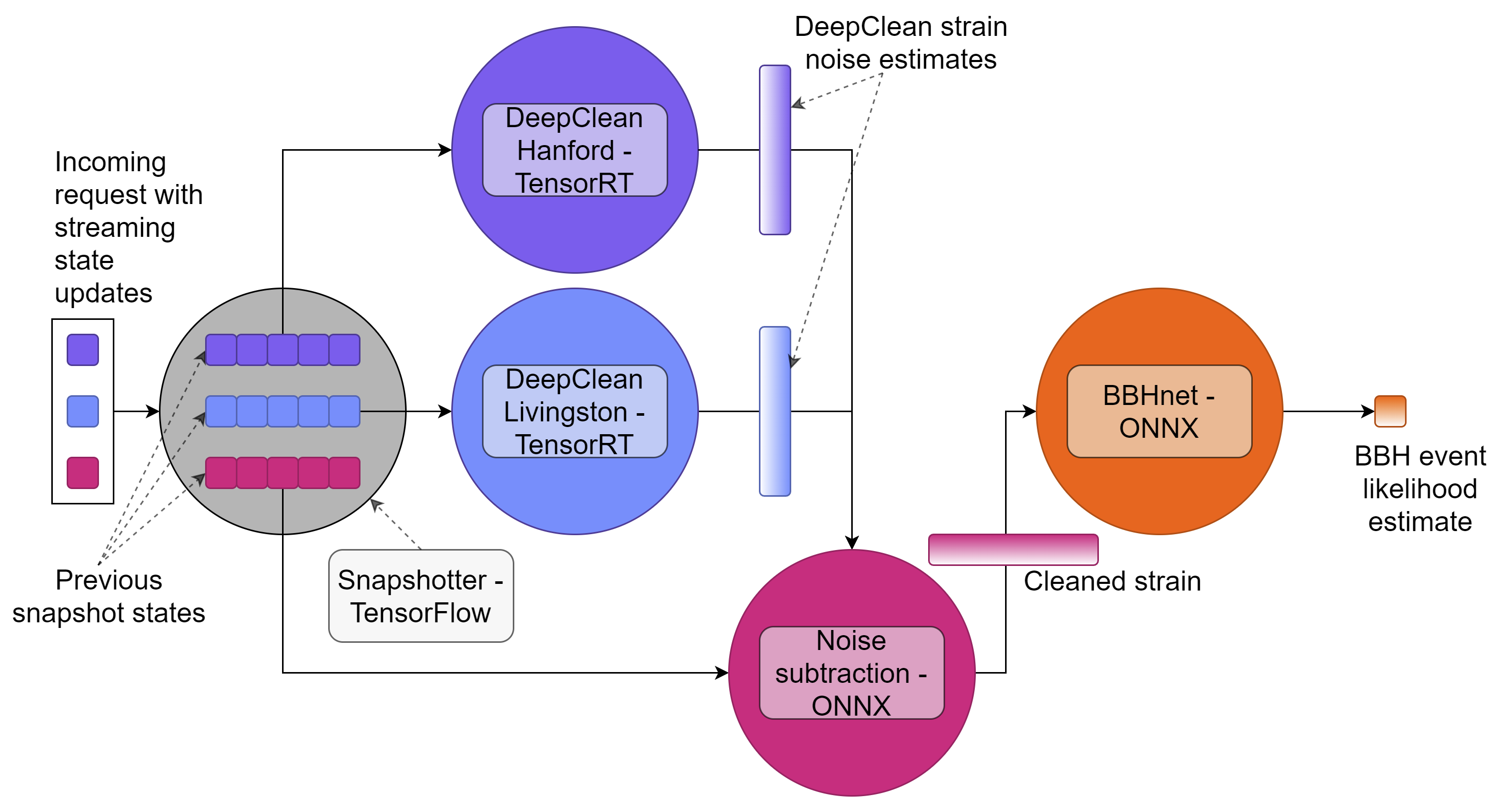}
\caption{Flow of data through ensemble of individual modules on inference service in the end-to-end pipeline. Snapshotter model maintains and updates state for all three data sources at once, then sends the updated snapshots to their respective downstream models. The framework backend used for each model is indicated in the legend. Using Triton's ensemble scheduler, arbitrary server-side pipelines like this can be constructed. The ability to construct complex ensembles of otherwise disparate models implemented using many software backends illustrates the robustness of the IaaS model.
}
\label{fig:aas_schematic}
\vspace{5mm}
\end{figure}

The second offline workflow, an end-to-end pipeline implemented in multiple different frameworks and comprises both DeepClean and BBHnet, is shown in Fig.~\ref{fig:aas_schematic}.

\begin{figure}[htb!]
\centering
    \includegraphics[scale=0.20]{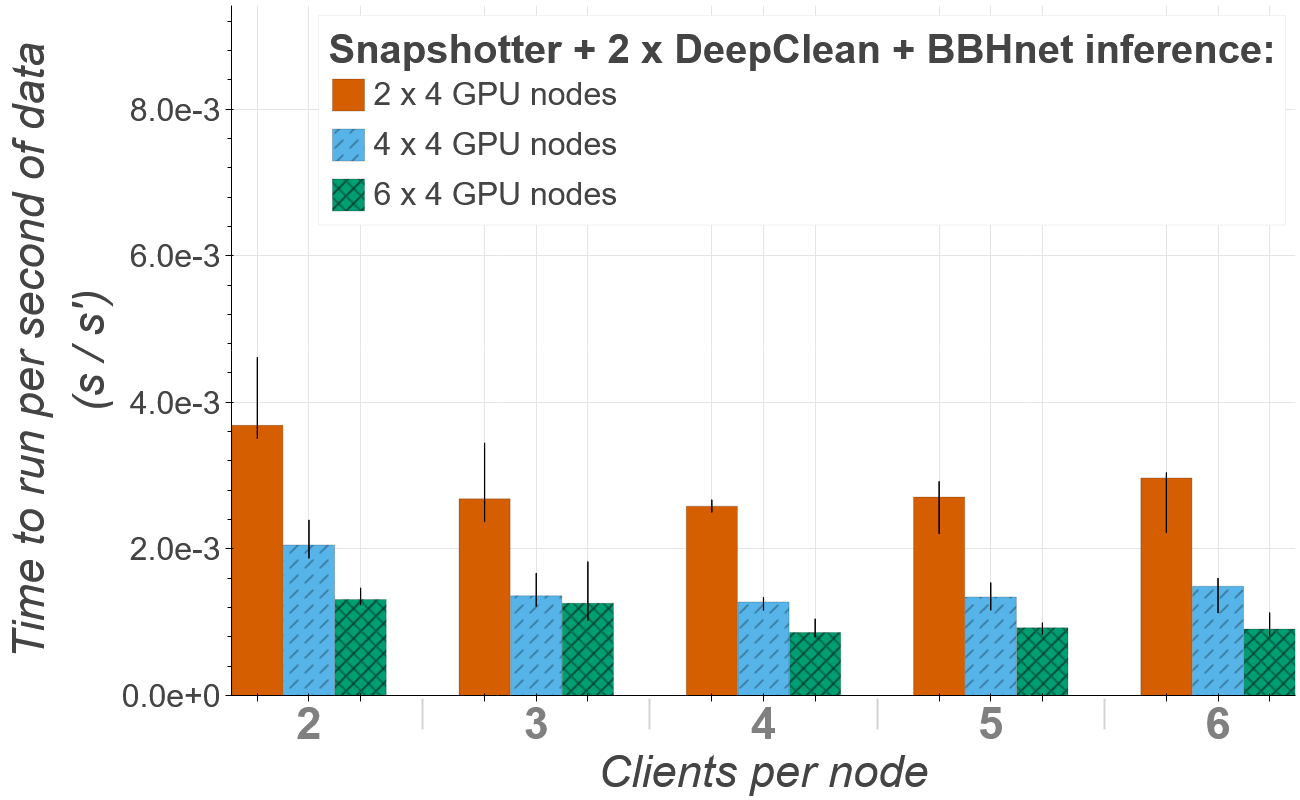}
    \caption{Distributions of time required to process one second of data for the end-to-end ensemble, $r = 10~\text{Hz}$. The colored bars represent the median value, with $\pm25$ percentiles depicted by the black bars. All nodes were equipped with 4 NVIDIA T4 GPUs and 32 vCPUs. The amount of concurrent execution per GPU was 6 for each DeepClean instance and 1 for all other models except the snapshotter.}
    \label{fig:e2e-time-r10}
\end{figure}

\begin{figure}[htb!]
\centering
    \includegraphics[scale=0.25]{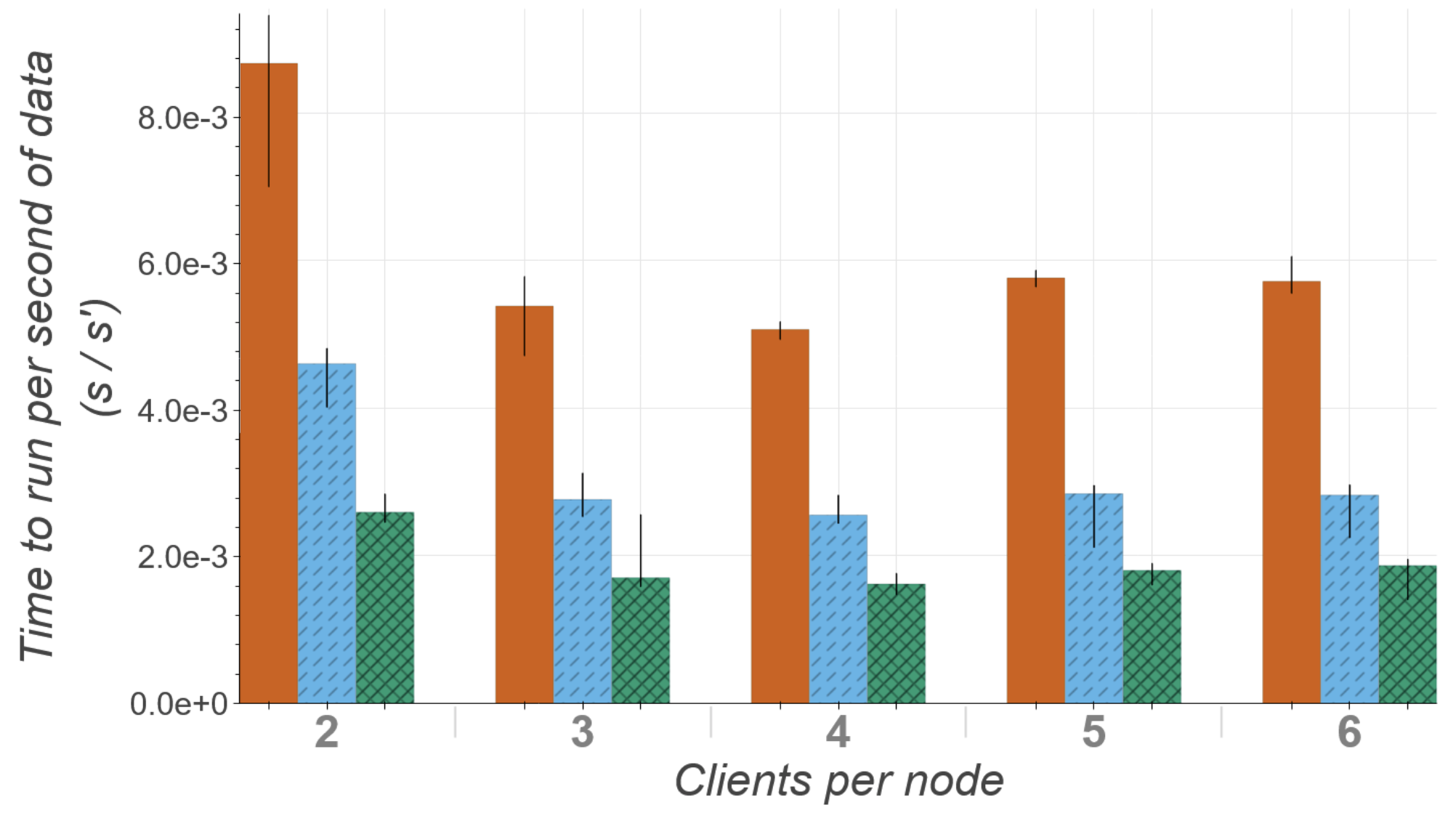}
    \caption{Distributions of time required to process one second of data for the end-to-end ensemble, $r = 20~\text{Hz}$. The colored bars represent the median value, with $\pm25$ percentiles depicted by the black bars. All nodes were equipped with 4 NVIDIA T4 GPUs and 32 vCPUs. The amount of concurrent execution per GPU was 6 for each DeepClean instance and 1 for all other models except the snapshotter.}
    \label{fig:e2e-time-r20}
\end{figure}

\begin{figure}[htb!]
\centering
    \includegraphics[scale=0.20]{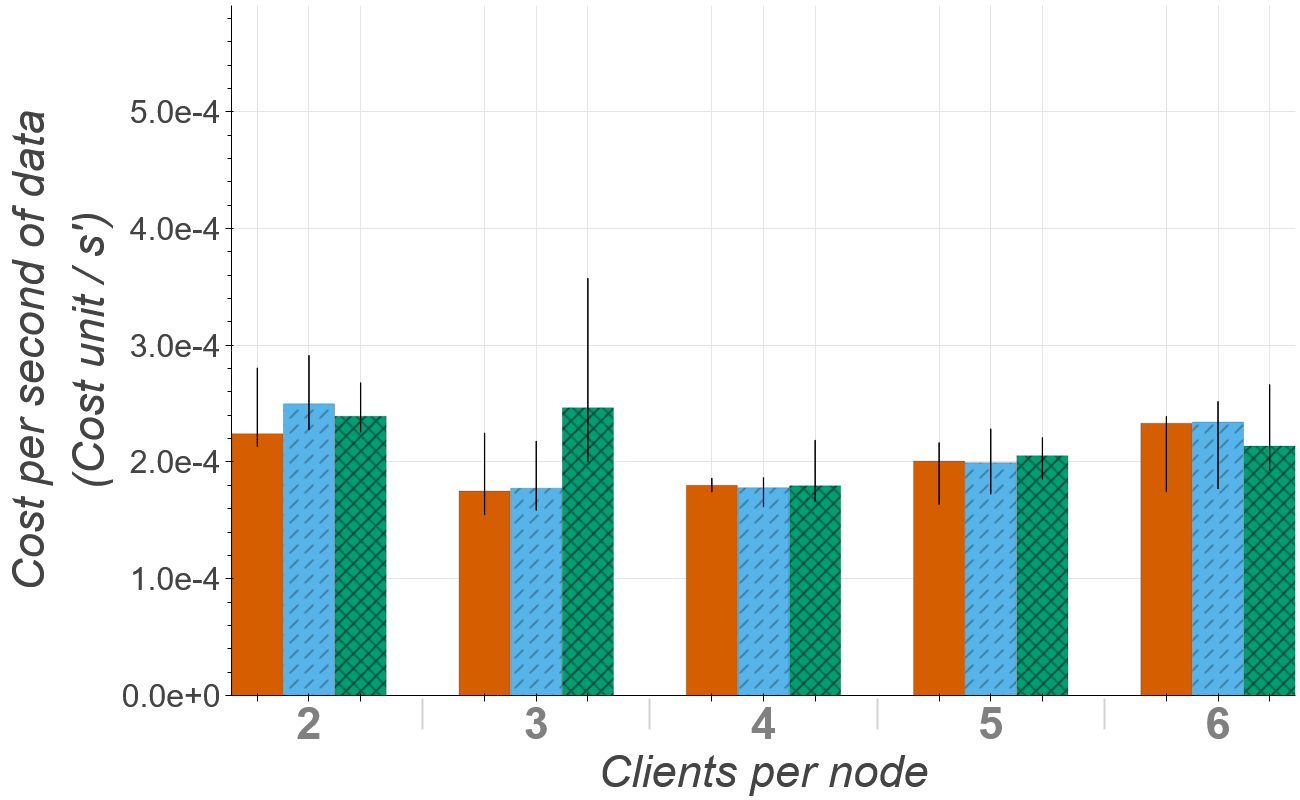}
    \caption{Distributions of cost required to process one second of data for the end-to-end ensemble, $r = 10~\text{Hz}$. The colored bars represent the median value, with $\pm25$ percentiles depicted by the black bars. All nodes were equipped with 4 NVIDIA T4 GPUs and 32 vCPUs. The amount of concurrent execution per GPU was 6 for each DeepClean instance and 1 for all other models except the snapshotter.}
    \label{fig:e2e-cost-r10}
\end{figure}

\begin{figure}[htb!]
\centering
    \includegraphics[scale=0.25]{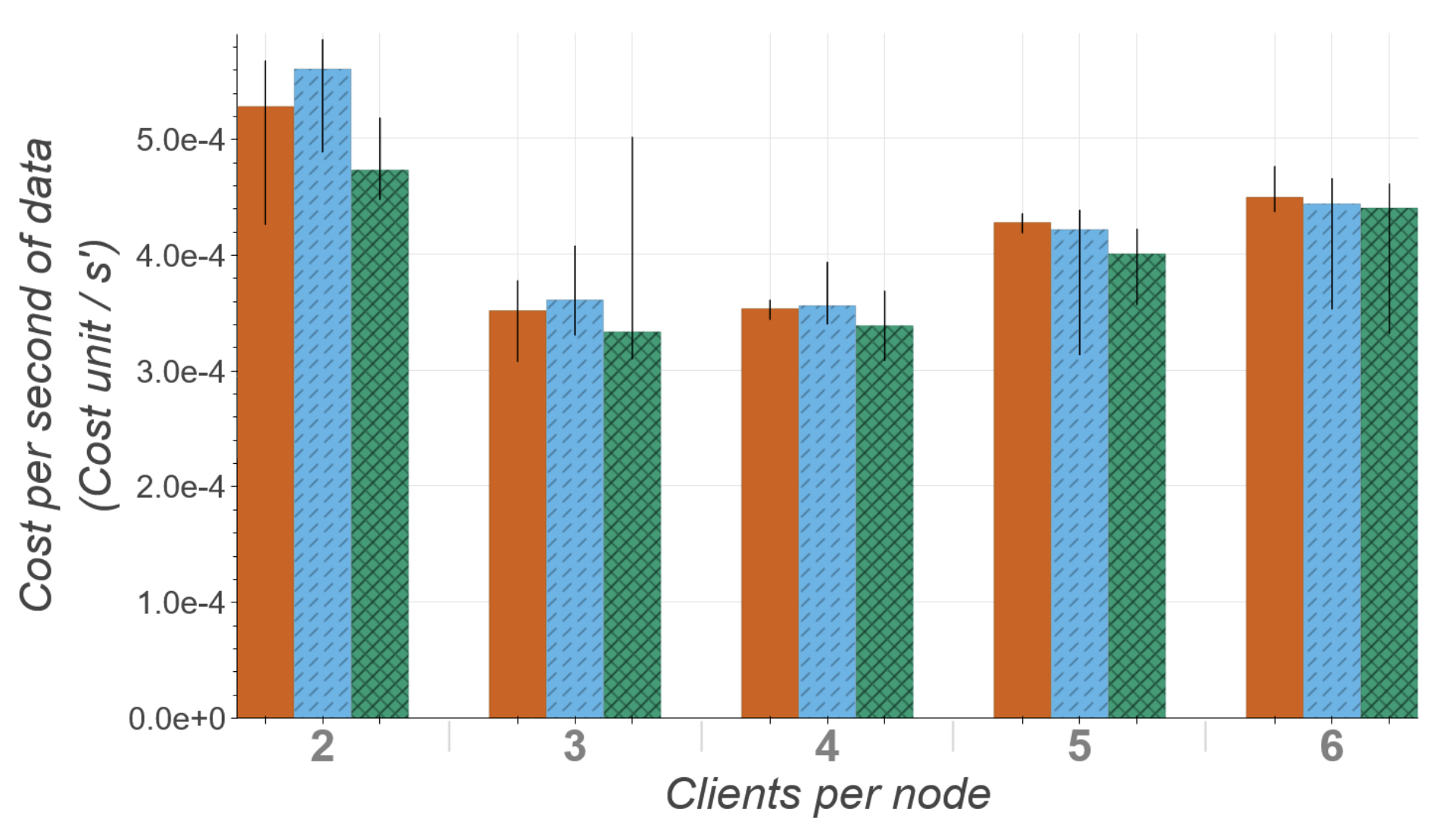}
    \caption{Distributions of cost required to process one second of data for the end-to-end ensemble, $r = 20~\text{Hz}$. The colored bars represent the median value, with $\pm25$ percentiles depicted by the black bars. All nodes were equipped with 4 NVIDIA T4 GPUs and 32 vCPUs. The amount of concurrent execution per GPU was 6 for each DeepClean instance and 1 for all other models except the snapshotter.}
    \label{fig:e2e-cost-r20}
\end{figure}

The time and cost required to process one second of data using this pipeline with various server and client configurations are shown in Figs.~\ref{fig:e2e-time-r10}~-~\ref{fig:e2e-cost-r20}.
The cost is computed by aggregating the cost-per-unit-time of all client and server resources and normalizing by the cost of 1 CPU hour.
See Methods for more information about the details of these measurements.
As can be seen from Figs.~\ref{fig:e2e-time-r10} and \ref{fig:e2e-time-r20}, for both values of the inference sampling rate ($r$) and the number of clients per server node, the processing time decreases nearly linearly as the number of server nodes is increased.
However, the total cost per second of data remains nearly constant with increasing numbers of nodes leveraged by the inference service, as shown in Figs.~\ref{fig:e2e-cost-r10} and \ref{fig:e2e-cost-r20}.
These trends show that once external constraints such as cost or sampling rate are imposed, the IaaS workflow is then able to make efficient use of the available resources, regardless of the exact values of the constraints.

We see in Figs.~\ref{fig:e2e-time-r10} and \ref{fig:e2e-time-r20} that increasing the clients per server node from 2 up to 4 is able to reduce the total processing time, regardless of the actual number of server nodes.
This reduction is possible because we do not fully utilize the server's resources, and therefore more clients can be served with the unused resources.
This underscores the ability of the IaaS paradigm to take full advantage of scarce server resources.  
For larger numbers of clients we still observe the same processing time, but an increased cost as a result of the saturated server throughput.
In this particular example, the models limiting the throughput are the two DeepClean instances, which are the larger of the models in the pipeline.
Some optimizations have already been applied to these models for this pipeline, but further improvement to their inference throughput is being investigated.

\begin{figure}[htb!]
\centering
\includegraphics[scale=0.3]{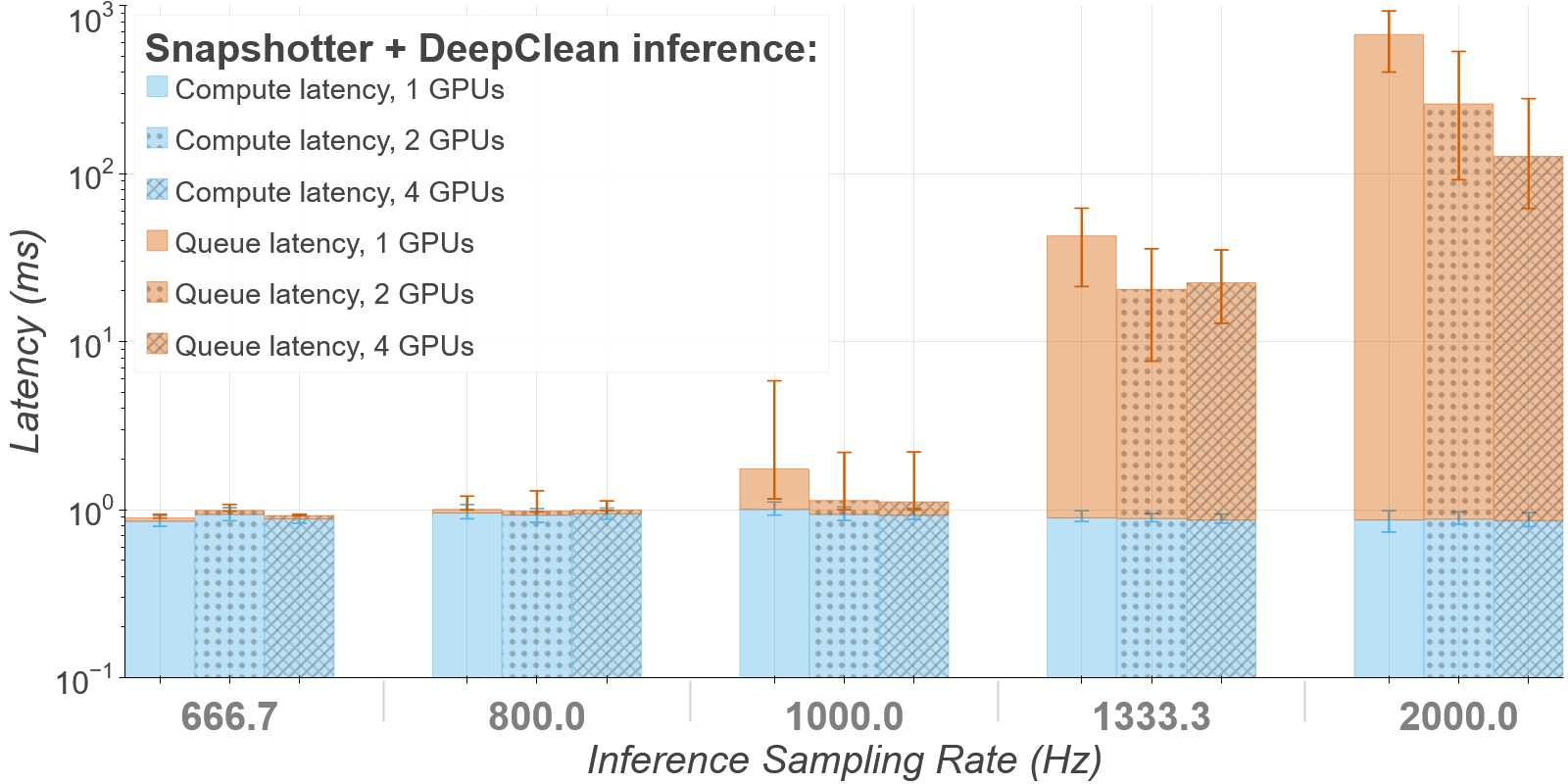}
\caption{End-to-end server-side latency for multiple GPU counts as a function of the inference sampling rate ($r$). Latency is broken down into contributions from both compute time and queuing time, with vertical bars representing median values and error bars representing $\pm25$ percentiles. The data sampling rate $f_s$ was fixed at 4000~Hz for all measurements, and all GPUs used were NVIDIA V100 16GB GPUs.
}
\label{fig:dc_online}
\vspace{5mm}
\end{figure}

The IaaS paradigm is equally capable of performing in an online setting.
Unlike in the offline settings described above, the online setting prioritizes low latency.
Figure~\ref{fig:dc_online} depicts the latency achieved with various server configurations as a function of the inference sampling rate $r$ for the DeepClean pipeline.
We disaggregate the latencies into time spent computing model inference (light blue) and time spent queueing for available resources (light brown).
At low values of $r$, the latency is dominated by compute time and remains nearly constant regardless of server configuration, since a single GPU is capable of handling the request load.
As $r$ increases past $\sim$1000~Hz, the request load overwhelms the maximal GPU performance, and so requests must queue and wait for resources to become available.
At the highest values of $r$, this resource availability is the primary determinant of total latency, which becomes a near-linear function of the amount of server resources.
More detailed inspection of latency sources indicates that the bottleneck in this pipeline is the streaming state update described in Methods, which limits the capacity for the downstream DeepClean model to benefit from additional GPUs.
Future optimizations to this update step via HHPC techniques will allow this pipeline to better utilize the available resources to both increase the values of $r$ which can be processed stably and decrease the latency incurred at sustainable values of $r$.

Finally, we perform tests emulating the offline end-to-end pipeline in the context of jobs running over a prolonged period of time.
The jobs are submitted to compute instances on the Google Cloud Platform via the HEPCloud framework~\citep{HoBa2017}. 
HEPCloud enables scientific experiments to access a heterogeneous variety of computing resources (on-premises batch, commercial cloud, supercomputing facilities), while presenting a single simple frontend interface (HTCondor) to the end user.
HEPCloud is a good test-bed for prototyping as-a-service workloads. 
First, it allows a large number of CPU, GPU, and other resources to be provisioned at a single site, facilitating scalable intra-site tests. 
Second, these resources form a self-contained system free from other jobs, which increases reproducibility for testing new frameworks. 
Last, the HTCondor frontend is widely used in pre-existing job submission frameworks.
The clients are configured to run on Google Cloud CPU nodes accessed via HEPCloud. 
The servers are configured to have between 4 and 80 NVIDIA T4 GPUs at the same site.

\begin{figure}[htb!]
\centering
\includegraphics[scale=0.3]{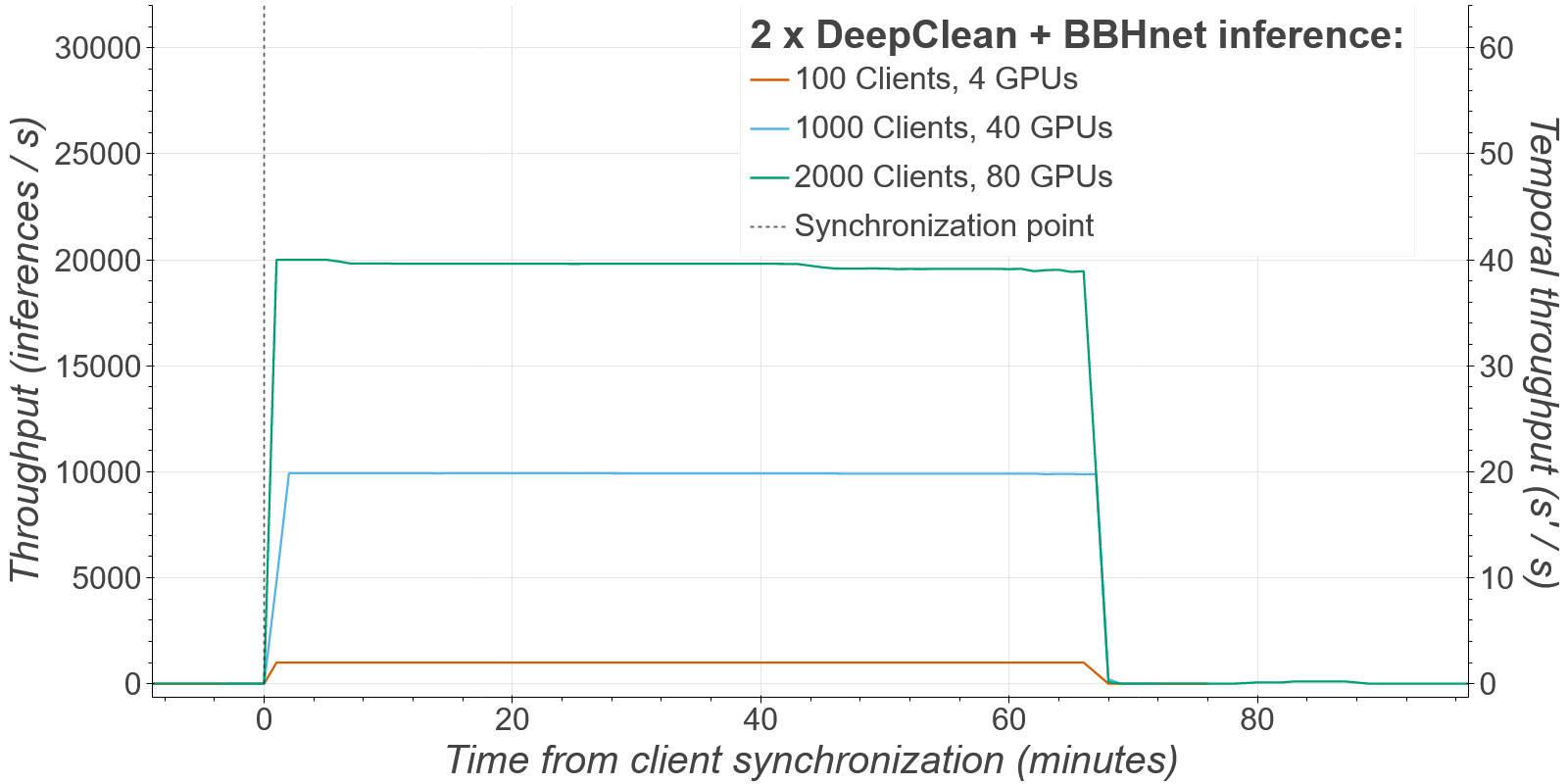}
\caption{The number of seconds of gravitational-wave data processed per second as a function of time for a sustained test using HEPCloud. The jobs are synchronized to start at the time indicated by the dotted vertical line.}
\label{fig:hepcloud_throughput}
\vspace{5mm}
\end{figure}

The result showing sustained throughput of frames (measured in seconds of data processed per second) as a function of time is shown in Fig.~\ref{fig:hepcloud_throughput}.
With a server consisting of 4 GPUs, stable processing of 1000 inferences per second (2 seconds of data per second) is observed.
This work is distributed across 100 HEPCloud clients. 
The observed throughput scales linearly with the number of servers (GPUs), reaching 20000 inferences per second (40 seconds of data per second) for 80 GPUs communicating with 2000 clients. 
This is a demonstration of our frameworks' ability to deliver inferences for a sustained period of time, and with large numbers of resources, using existing gravitational-wave experimental paradigms.

\begin{figure}[htb!]
\centering
\includegraphics[scale=0.75]{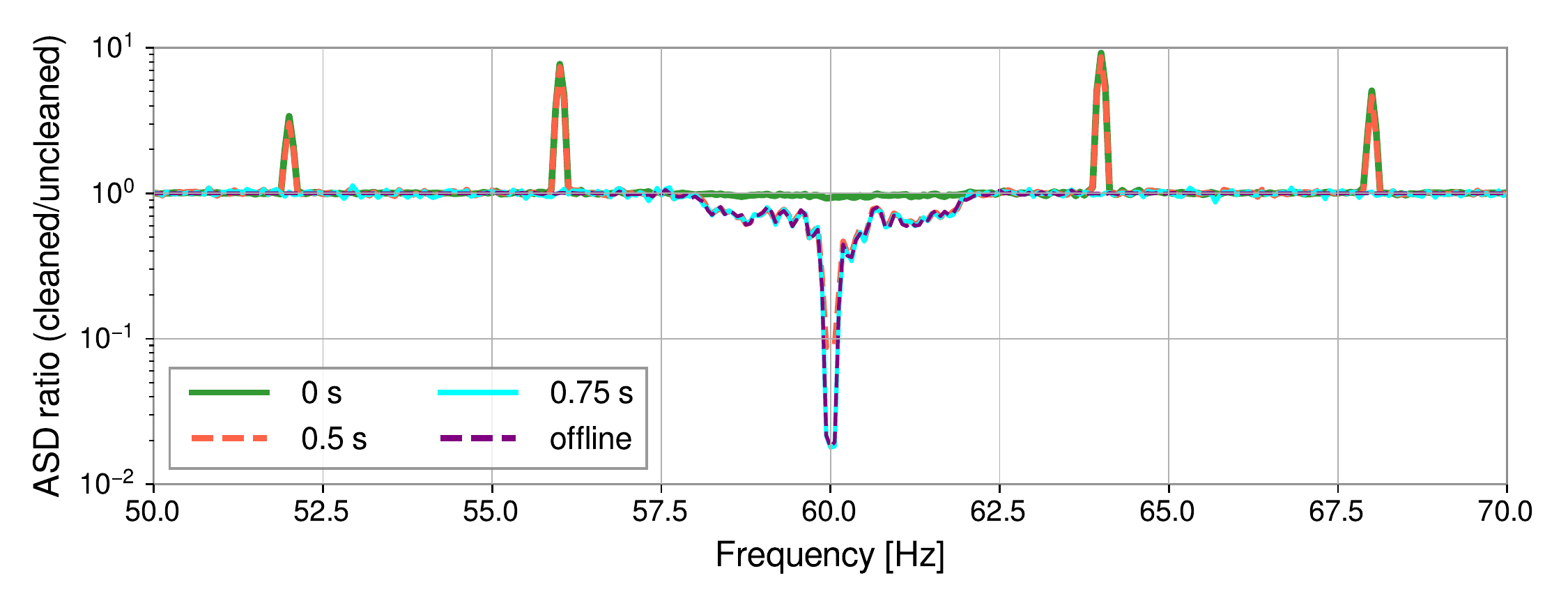}
\caption{Performance of the online-deployed DeepClean noise regression as a function of the {\it aggregation latency} - length of the time series data excluded from each cleaned segment, to avoid noisy edges. The quantity on the y-axis is the ratio of the Amplitude Spectral Density (ASD) of the cleaned data to that of the uncleaned data. We compare the offline case (purple) to an analysis with zero latency (green), 0.5 s (orange), and 0.75 s latency (cyan). At zero latency, the fraction of frequency bins within the [50,70] Hz that differ from the offline ASD ratio by more than 10\% is  $\sim$23\%; this quantity reduces to $\sim$4\% and $\sim$1\% with 0.5 s latency and 0.75 s latency respectively. 
}
\label{fig:dc_performance}
\vspace{5mm}
\end{figure}

While we have shown that the IaaS paradigm is capable of meeting the computational needs of streaming low-latency data denoising and astrophysical searches, there are additional considerations that are required for a fully real-time pipeline for multi-messenger followup. 
Specifically, the performance of the deep neural network inference pipeline must operate with the same fidelity in the offline scenario. 
In this instance, we have observed that there is some degradation in the subtraction quality at the edges of the cleaned segments when using DeepClean.  
While we leave the mitigation of this effect to future work, in the present setup we simply exclude the quality-degraded edges from the cleaned data segments at a latency cost equal to the excluded data length.
We refer to this latency as the aggregation latency.
Fig.~\ref{fig:dc_performance} demonstrates this degradation and subsequent recovery by comparing the performance with the fully offline DeepClean pipeline.
We see that an aggregation latency of 0.75 s is able to closely reproduce the amplitude spectral density of offline DeepClean.
While this limits the minimum possible latency for the online pipeline we have used, it could potentially be reduced or even removed entirely by an algorithm designed (trained) specifically for low-latency cleaning.

Another factor which must be considered when discussing the overall latency is how often the trained network must be updated on the server. 
Typically a cleaning algorithm must be retrained on recent data to maintain performance under gradually changing conditions.
For continuous functioning, the trained model must remain valid for longer than the time it takes to retrain a new model. 
For DeepClean, training the network on a new dataset typically takes $<$20 minutes on a single GPU, while analysis of O3 gravitational-wave data from the LIGO-Hanford detector shows that a trained model is effective for at least 4096\,s of subsequent data, a common storage length for gravitational-wave data.
Therefore, in the case of DeepClean there is sufficient time for an update cycle.

We have demonstrated a fully realistic computational workflow to process gravitational-wave data with ML algorithms using a heterogeneous computing stack.
Our workflow comprises both an ML-based denoising algorithm and an ML-based binary black hole detection algorithm.
By caching the time-series, we can increase the input throughput by several orders of magnitude.
Our workflow can seamlessly incorporate future updates to either algorithm and can be extended to additional algorithms for detection or any other analysis.
We have run this workflow with real gravitational-wave data and demonstrated operation in two different scenarios: online and offline data processing. 
We find that in the online scenario our current setup can perform real-time noise subtraction and binary black hole detection with a latency of one second, and is currently limited by the performance of the noise cleaning algorithm itself, and not the inference latency. 
In spite of this limitation, a server with a GPU located locally at each gravitational-wave site would be able to output a cleaned stream of data within one second of acquisition.
With modified deep neural network models, it is likely that the latency can be reduced to milliseconds.
The success of this scheme is not dependent on the specifics of the algorithms implemented, which is crucial for its long-term viability as computational needs and resources develop.

For offline gravitational-wave data processing, we have set up a full reprocessing stream.
By relying on the Inference as-a-Service model, we are able to optimally configure the GPUs and CPUs to process the data-stream, leading to large increases in the overall throughput of the system. 
Depending on the desired hardware setup, we demonstrate orders of magnitude reductions in the time required to process gravitational-wave data. 
Our system is dynamically scalable, and remains optimized whether we use a small number of GPUs or a larger number of resources. 
We have also demonstrated scalability by testing a sustained offline workflow with HEPCloud on Google Cloud. 
As a consequence, we have demonstrated that we can scale out gravitational-wave reprocessing to utilize a large amount of computing nodes available within a cloud or high performance computing center.
With sufficient computing resources, our computing scheme can reprocess gravitational-wave datasets containing many years of data within a few hours. 

To conclude, we have demonstrated a fully realistic integration of ML-based hetereogeneous computing within existing gravitational-wave computing stacks that are ready to be deployed.
Our implementation shows the ability to meet latency and scale requirements for online and offline uses in gravitational-wave ground-based interferometers like LIGO, Virgo and KAGRA.
It also presents the computational platform for incorporating ML-based techniques for real-time controls of the numerous servo loops that make part of the laser interferometry in present and future ground interferometers~\citep{reitze2019cosmic} and statisfies the requirement for sub-ms latencies and ability to handle thousands of channels anywhere in the 1-10MBps bandwidth needs~\citep{Abbott_2016} have already been achieved with off-the-shelf computing hardware our implementation uses.
This work also has broad implications across many fields where the processing of real-time data-streams is critical, including areas of electromagnetic and neutrino astronomy. 
Most importantly, this work can significantly improve our ability to perform multi-messenger astronomical observations. 
As gravitational-wave detectors become increasingly sensitive over the course of second-generation improvements in this decade~\citep{Abbott_2020_obs}, and with third-generation improvements in the next~\citep{reitze2019cosmic}, this new heterogeneous computational stack has the ability to facilitate the computational demands needed to accelerate discovery.

\newpage

%TC:endignore

%\clearpage
%\bibliographystyle{naturemag}
%\bibliography{references}

\section{Acknowledgements}
The authors are grateful for computational resources provided by the LIGO Laboratory at Caltech, Livingston, LA and Hanford, WA. The LIGO Laboratory has been supported under National Science Foundation Grants PHY-0757058 and PHY-0823459.
AG, DR, TN, PH and EK are supported by NSF
grants \#1934700, \#1931469, and DR additionally with the IRIS-HEP grant \#1836650. JK is supported by NSF grant \#190444.
SM and MC are supported by NSF grant PHY-2010970. Work supported by the Fermi National Accelerator Laboratory, managed and operated by Fermi Research Alliance, LLC under Contract No. DE-AC02-07CH11359 with the U.S. Department of Energy. The U.S. Government retains and the publisher, by accepting the article for publication, acknowledges that the U.S. Government retains a non-exclusive, paid-up, irrevocable, world-wide license to publish or reproduce the published form of this manuscript, or allow others to do so, for U.S. Government purposes.
Cloud credits for this study were provided by Internet2 managed Exploring Cloud to accelerate Science (NSF grant PHY-190444). Additionally we would like to thank NSF Institute for AI and Fundamental Interactions (Cooperative Agreement PHY-2019786).
The authors are also grateful for the support provided by Stuart Anderson in the realization and testing of our workflow within the LDG.
Finally, the authors would like to thank Alexander Pace for providing useful comments on the manuscript.

The authors declare that they have no
competing financial interests.

AG and DR are the primary authors of the manuscript. JK integrated applications in HEPCloud. ST and BH support and operate HEPCloud. SM, MC, EK, and TN support development of DeepClean. All authors contributed to edits to the manuscript.

Correspondence and requests for materials
should be addressed to Alec Gunny~(email: alecg@mit.edu).

\clearpage
\newpage

\appendix
\section{Gravitational Wave Computing Architecture}

Our work builds off previous efforts in the realms of as-a-service computing and ML.
Significant motivation is taken from the integration of these concepts for usage in high energy physics (HEP) where recent algorithmic advances and the availability of large datasets have greatly facilitated the adoption of ML.
Previous work with experiments at the CERN Large Hadron Collider and the ProtoDUNE-SP experiment at the Fermi National Accelerator Laboratory have shown that the as-a-service computing model has the potential to offer impressive speed-ups, improved performance, and reduced complexity relative to traditional computing models~\citep{Krupa:2020bwg, Rankin2020FPGAsasaServiceT, Wang:2020fjr}.
These works have also demonstrated the ability to perform inference as-a-service with both GPUs as well as FPGAs.

While promising, this paradigm offers several challenges from an engineering perspective. To ensure consistent deployment, as is the case with rule based algorithms, applications in this model (trained neural networks) need to be versioned, tested, and validated.
They may be asked to share pools of compute resources, including comparatively expensive co-processors like GPUs or FPGAs, with other applications, or may require deployment on dedicated hardware to meet strict constraints on inference latency.
Moreover, these applications will need to be shared among users who may not be knowledgeable or interested in the details of their implementation (save for some guarantees about their functionality).
The pipelines built on top of them will need to be kept up-to-date as newer versions are fine-tuned on new data or replaced wholesale by novel architectures, and frequent updates may impose a non-trivial burden on users to avoid degraded or interrupted service for sensitive applications.

The computing architecture traditionally employed in managing gravitational-wave searches is based on the HTCondor batch system~\citep{DBLP:journals/corr/abs-2011-14995}.
The individual tasks that make up an astrophysical search pipeline are performed by modules which have been programmed and optimized by data analysis experts specific for each search.
The resulting workflow is often made up of (often tens of) thousands of jobs that require little or no communication amongst them, thus allowing their straightforward parallelization.
HTCondor provides the mapping of such jobs to resources and shepherds them to completion.
This computing model requires data to be discoverable either through bulk, local storage available on each compute site or via migration of data to the site where jobs are assigned, e.g., when using Open Science Grid resources~\citep{Pordes_2008}
 (\url{https://opensciencegrid.org/}). 

\section{Gravitational-wave data processing}

\subsection{Streaming Iaas}
One feature of gravitational-wave data that makes the application of the IaaS model more challenging is that the data of interest comprises streaming time series that need to be processed sequentially, which limits the extent to which an inference service can leverage scale to process data in parallel.
This is particularly true for convolutional models like DeepClean and BBHNet, which process fixed-length ``snapshots'' of this time series, sampled at some frequency $r \leq f_s$, where $f_s$ is the sampling rate of the time series.

\begin{figure}[htbp!]
    \centering
    \includegraphics[scale=0.5]{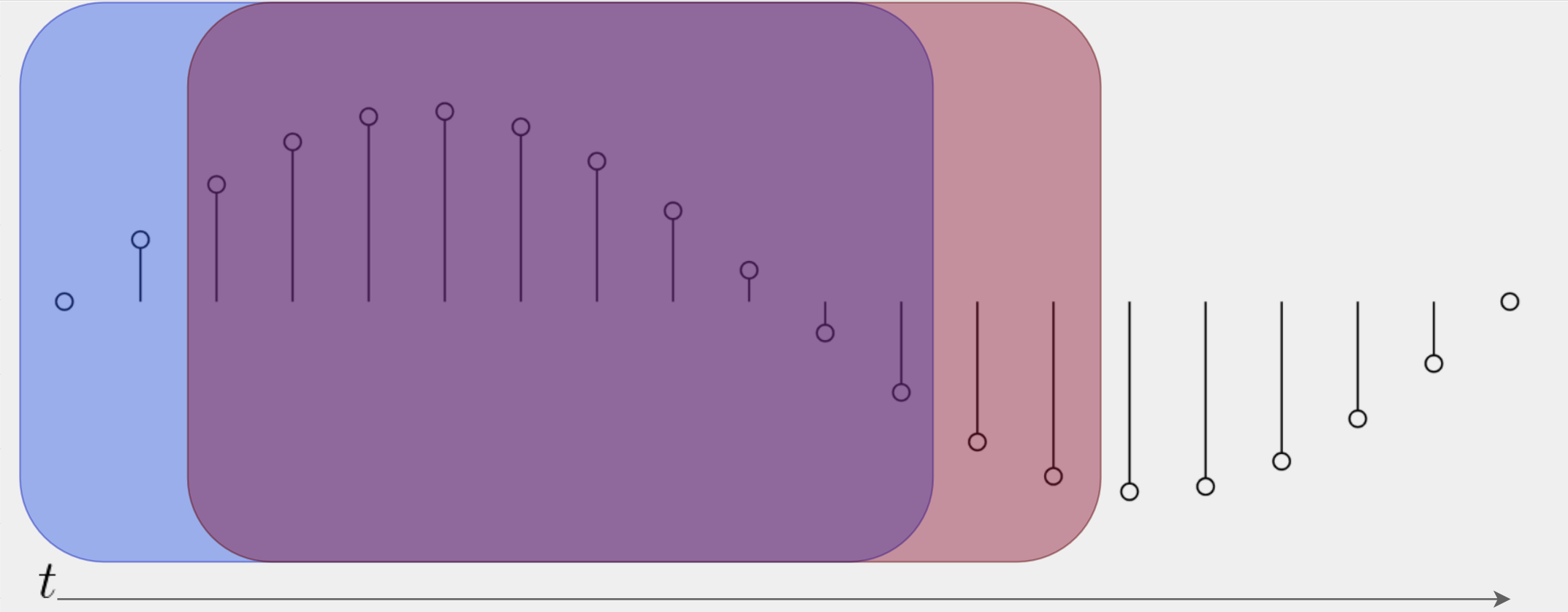}
    \caption{Overview of snapshotter behavior, first step. High frequency inference relative to the length of the frame results in redundant input data. Colored squares represent time series snapshots at different points in time.}
    \label{fig:snapshotter_overlap}
\end{figure}

As shown in Fig.~\ref{fig:snapshotter_overlap}, higher values of $r$ compared to the inverse of the length of the frame will produce highly redundant data between subsequent frames.
If requests made to ``static'' models like DeepClean or BBHnet provide a full snapshot's worth of data as input, their network load will increase substantially, bottlenecking the pipeline.

\begin{figure}[htbp!]
    \centering
    \includegraphics[scale=0.7]{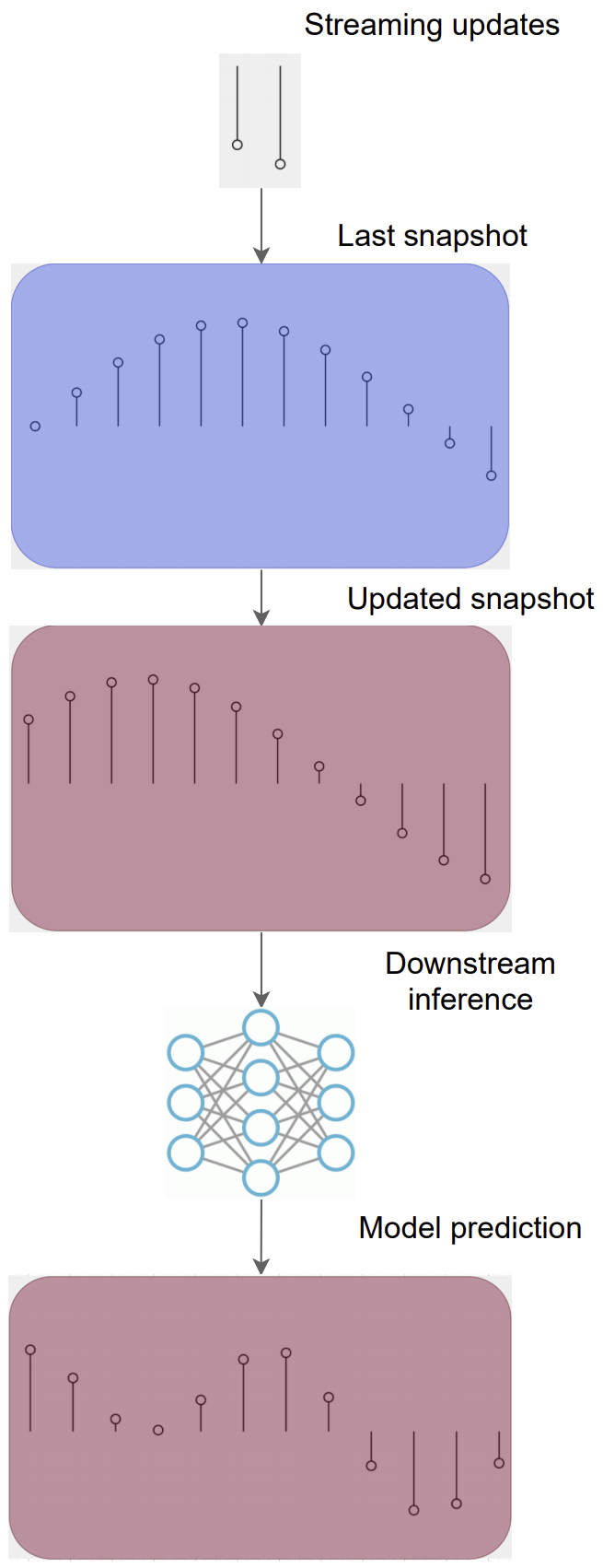}
    \caption{Overview of snapshotter behavior, second step. Streaming new samples to the snapshotter updates its state, which is then passed to downstream networks for inference. (c) Fully-online inference for models that produce time series output streams back the last $\frac{f_s}{r}$ samples at each timestep. Outputs aren't identical between frames, and there is potential for improved performance by aggregating overlapping predictions.}
    \label{fig:snapshotter_action}
\end{figure}

\begin{figure}[htbp!]
    \centering
    \includegraphics[scale=0.7]{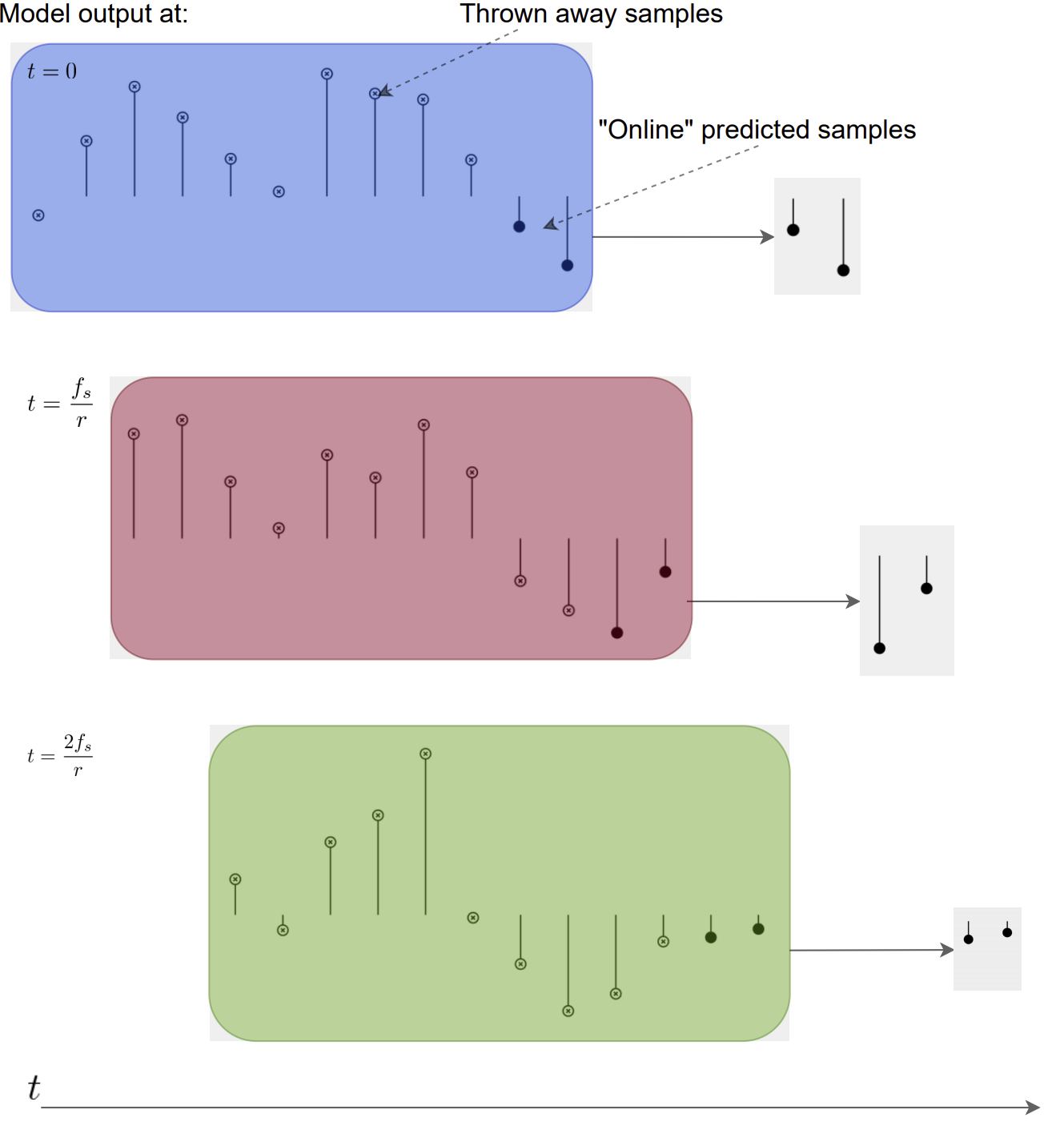}
    \caption{Overview of snapshotter behavior, third step. Fully-online inference for models that produce time series output streams back the last $\frac{f_s}{r}$ samples at each timestep. Outputs aren't identical between frames, and there is potential for improved performance by aggregating overlapping predictions.}
    \label{fig:snapshotter_output}
\end{figure}

To get around both of these limitations, we host an extra model on the server which maintains a state representing the previous input snapshot.
Inference requests are made to this model  providing a $\big{\lfloor}\frac{f_s}{r}\big{\rfloor}$-sized stream of strictly new samples that are used to update the state to a new snapshot, which is then passed to downstream static models. By providing only the new time series a dramatic reduction in networking bandwidth is needed equal to the sampling rate divided by the total model input size; for the models used in this paper, this reduction is more than 1000-fold. 
A diagram of this data flow is shown in Fig.~\ref{fig:snapshotter_action}.

While updates to the snapshot state have to be performed sequentially, downstream inference on adjacent frames can be performed in parallel with a server node by leveraging Triton's ensemble scheduling API.
In the online setting, this means that as long as the state update can be performed faster than $\frac{1}{r}$ seconds, we avoid introducing a bottleneck into the pipeline.
In the offline setting, we can host multiple snapshot instances on a single node and parallelize inference across them in order to saturate the downstream throughput capacity.
This ability to cache states in order to better utilize complex ensembles of downstream models represents a significant optimization that is critical to enabling the effective use of deep learning inference in gravitational-wave physics and astronomy.
In all experiments described below, the snapshotter model was implemented using TensorFlow~\citep{tensorflow2015-whitepaper}.

\subsection{Data aggregation}

There remains one additional challenge in the online streaming setting for architectures like DeepClean whose output is itself a time series with a data sampling rate greater than $r$.
For these models, there will also be overlap between the \textit{output} segments produced by each inference.
Unlike the input redundancy, however, the overlapping segments will not be identical: each prediction will be conditioned on data from different windows in time and therefore will, in general, be distinct.

While one may improve prediction quality by aggregating the overlapping segments from subsequent windows, this would incur additional latency in the online setting, as completed predictions would have to wait for inference on subsequent frames to complete in order to aggregate them.
For this reason, we have chosen to adopt the ``fully-online'' or ``causal'' prediction scheme shown in Fig.~\ref{fig:snapshotter_output}, in which predicted segments keep only their most recent $\big{\lfloor}\frac{f_s}{r}\big{\rfloor}$ samples.

\subsection{Aggregation latency}

As discussed in the Main Text, DeepClean subtraction quality degrades at the edges of cleaned segments for a span of data with length $\lesssim1\,s$. For the offline analysis, we typically perform cleaning on 4096\,s of data, i.e only $\sim$0.02\% of the data is impacted. However, for the online analysis where data is cleaned in real-time, we use shorter data segments ($\sim$\,8\,s), increasing the importance of problematic segment edges. We currently address this as follows, although future work will likely improve upon this technique. In order to clean the most recent $1\, s$ data, {\it i.e,}  a window of  $[-1\,s$, \, $0\,s$], if we take the noisy edge to be $\delta t\,s$ and analyze an $8\,s$ window; $[-8+\delta t\, s$, \, $0+\delta t\, s]$, we can safely recover the cleaned sub-window $[-1,0]\,s$ after removing the $\delta t\,s$ of noise-affected subtraction. This increases our latency by $\delta t\,s$; we refer to this as our {\it aggregation latency}, as it is due to how our cleaned data is aggregated.  

Fig.~\ref{fig:dc_performance} shows the performance of the subtraction as a function of the {\it aggregation latency}.
Here, we have taken, as an example, the subtraction of the 60\,Hz power-line and its harmonics in O3 data from the LIGO-Hanford detector. 
The plot shows the ratio of the Amplitude Spectral Density (ASD) of the cleaned data to the original data as a function of frequency. In the ideal scenario, only 60\,Hz noise is removed, leading to  a
dip in the ASD ratio near 60\,Hz while it is $\sim 1$ elsewhere. This is true for the offline analysis case (purple) with a dip of two orders of magnitude. With zero latency (where the noisy edge is fully included), no subtraction was achieved and some additional noise from aggregation was added to the data, appearing as upward excursions in the ASD ratio. With an aggregation-latency of 0.5\,s, similar features are observed, except for some subtraction at 60\,Hz. With 0.75\,s latency, the offline subtraction is reproduced, meaning that cleaned data can be delivered only 0.75\,s after the original data. Our tuning has shown that aggregation latency can be controlled by tuning the duration of data per inference and the overlap between successive inference segments. In addition, the network loss function likely affects this. Unlike the current models that are optimal for cleaning bulk offline data, the objective function may need to be changed to reflect the online requirements, to improve performance. 

\subsection{Online Deployment}

As noted above, the online deployment scenario more generally is characterized by a sensitivity to the latency incurred by any individual request.
It is not sufficient that \textit{all} requests be completed by some deadline; it is also required that, with high probability, each request (or subset of requests) be completed within some fixed amount of time after it was generated.
In this setting, an operating point can be identified by the amount of compute resources being used by the service and the level of parallelism being leveraged on those resources via concurrent execution of inference.
Each operating point is associated with some cost as well as distributions of the latency and throughput those resources can achieve.

The online gravitational-wave data use-case studied in this paper subtracts noise from strain data in real-time using DeepClean, increasing the sensitivity of downstream event detection algorithms.
It is constrained by the exact mechanism and overall organization by which gravitational-wave detectors produce input data.
The nominal data acquisition model in gravitational-wave experiments relies on front-end computers that accumulate 1/16th of a second of data that are sent in chunks to a data aggregator on a fast network~\citep{BORK2021100619}.
The lowest latency data this aggregator makes available for analysis correspond to 1-second long gravitational-wave ``frames'' -- the standard format for storing gravitational-wave data~\citep{T970130} -- that are compiled in memory and are made available for prompt analyses.
In the future, we hope to be able to move our pipeline closer to the front-end data source in order to avoid the latency incurred by waiting for full frames to aggregate.

In this setting, throughput is not a dependent variable to be measured, but rather an independent variable that is fixed by $r$.
Sustaining a given throughput level means either scaling up resource usage in order to achieve lower latency levels, or increasing the level of parallelism on a smaller resource pool at the expense of higher latency.
Fig.~\ref{fig:dc_online} measures this trade-off at incremental values of $r$, using 21 witness channels and $f_s = 4000~\text{Hz}$ for all cases.
As mentioned in the main text, the experiments at $r = \frac{f_s}{3}$ and $r = \frac{f_s}{2}$ were already bottlenecked by the state update at the snapshotter, so measurements were not attempted for $r = f_s$.
The amount of parallel execution was searched over at each value of $r$, and the values reported in Fig.~\ref{fig:dc_online} represent measurements at the level of parallelism that incurred the least median latency.
These measurements were made using NVIDIA V100 16GB GPUs, which were connected directly via an NVLink connection for the multi-GPU cases.
The software backend used to implement DeepClean was TensorRT using FP16 precision.

The pipeline for these measurements was deployed entirely on LDG resources in order to minimize data transfer latencies.
The client pipeline loads and preprocesses frames written to LDG shared memory by a data replay service.
It then asynchronously packages $\big{\lfloor}\frac{f_s}{r}\big{\rfloor}$-sized chunks of these frames into requests to stream to the inference service.
Responses are compiled asynchronously until a full frame’s worth of noise estimates is constructed.
This frame is postprocessed then subtracted from the associated strain data, which is then written to disk.
The latency samples depicted in Fig.~\ref{fig:dc_online} represent measurements of the time taken to execute one full inference on the server side.
Each sample is  the average value over a $\sim$50~ms interval, as reported by Triton's metrics service.

\subsection{Offline Deployment}
Offline inference scenarios, on the other hand, are marked by their insensitivity to latency.
Because any post-hoc analysis tends to be done on the inference results in bulk, it matters less how long any \textit{particular} request takes to complete.
As such, the relevant metrics pertain to how \textit{quickly} data can be processed in aggregate, i.e. throughput, and how much that speed \textit{costs}.
Moreover, because all of the data in the offline scenario already exists up front, $r$ has no relationship to the rate at which data can be generated, and instead dictates the \textit{number} of inferences that need to occur for any given length of data.

For our offline pipelines, we split the datasets into segments and distributed these segments evenly among many client instances, whose requests are then distributed evenly among the available inference service instances.
The number of snapshotter instances per GPU on each service instance is scaled to meet the number of clients assigned to that instance.
Fig.~\ref{fig:triton-cloud} roughly depicts the network diagram for both offline use-cases studied here.
Client VMs read frame files from a regionally co-located cloud storage bucket and resample them to $f_s = 4096~\text{Hz}$ before applying preprocessing and streaming requests to the inference service.
Service instances are deployed in the same regional datacenter using Google Kubernetes Engine.
Each DeepClean instance described below uses 21 witness channels.

The measurements depicted in Fig.~\ref{fig:dc_offline} began as throughput estimates over $\sim$50~ms intervals by counting the number of inferences that took place during each interval as reported by Triton’s metrics service.
The $r$ value for each experiment was divided by these throughput samples to produce samples in units of seconds per second of data, from which the distributions in Fig.~\ref{fig:dc_offline} were formed.

% used times 1238166016-1240758272 GPS time
The first offline use-case described above used DeepClean to subtract noise from roughly one month of gravitational-wave strain data collected during the O3 run, now contained in $\sim$10,000 frame files of length 256 seconds each.
The distributions of processing time per second of data for this pipeline are depicted in Fig.~\ref{fig:dc_offline}, with $r = 0.25~\text{Hz}$ for all measurements.
For the LDG pipeline, samples were taken directly by measuring the time delta over multiple processing runs.
The GPU used in this pipeline was an NVIDIA V100 16GB GPU.
For all the IaaS experiments, each service is associated with 6 client instances, and inference is executed for DeepClean using the ONNX runtime~\citep{bai2019}.
The CPU-only IaaS pipeline utilized 64 vCPUs and 6 concurrent DeepClean execution instances, while each GPU-enabled node utilized 32 vCPUs and 6 concurrent execution instances per GPU.
We leveraged NVIDIA V100 32GB GPUs on all GPU-enabled inference service nodes in order to maintain consistency with the resources available on LDG, but similar experiments leveraging NVIDIA T4 GPUs showed comparable throughput at a much lower price point and would likely be used in production.
For all the IaaS pipelines, each client aggregates responses and produces clean data in much the same way the online DeepClean pipeline does, with cleaned strain data written to another cloud storage bucket.

The second offline use-case performed inference using an end-to-end ensemble depicted in Fig.~\ref{fig:aas_schematic}, which also specifies the software framework used to execute each model.
This pipeline uses two DeepClean instances to remove noise from strain data from \textit{both} detectors, which are combined and postprocessed then passed to BBHnet to produce event likelihood estimates.
% used times 1185402880-1185501184 GPS time
It was run on $\sim$27~hours of data collected during the O2 run.
Each inference service node in the pipeline leveraged 4 NVIDIA T4 GPUs and 32 vCPUs, with 6 concurrent execution instances available per GPU for each DeepClean model and 1 for all others except the snapshotter model.
The time and cost per second of data distributions for this pipeline are depicted in figs. \ref{fig:e2e-time-r10}-\ref{fig:e2e-cost-r20}.
The variation in Fig.~\ref{fig:e2e-cost-r20} at 6 nodes with 3 clients per node is due to implementation issues that caused clients to fail to synchronize appropriately, and does not reflect volatility in the inference service's throughput capacity.

\subsection{HEPCloud}

We use the HEPCloud~\citep{HoBa2017} package to submit jobs which run client pipelines on virtual nodes on Google Cloud. 
The jobs are submitted using HTCondor from the Fermilab computing cluster. 
The Google Cloud nodes running the clients are configured as to have 16 virtual CPUs and 30 GB of memory. 
The Triton server containing the ML model workflow is deployed on Google Cloud: each server node is configured to have with 64 vCPUs, 240 GB memory, and four NVIDIA T4 GPUs. 
We test servers configured to have 4, 40, and 80 GPUs. 
A non-streaming version of the end-to-end ensemble is employed. 
The model pipeline therefore consists of the DeepClean and BBHnet models, which are hosted on a bucket on Google Cloud Storage.
For optimal throughput, each GPU is configured to host six instances of the DeepClean models and one instance of the BBHnet model. 
The IP address belonging to each server is passed to a defined set of clients.
For the purposes of this experiment, a client:server ratio of 100:1 is found to offer sustained throughput. 
Each client is allocated more than an hour of data to process.  
To ensure consistency and reproducibility in testing the workflow, the clients are synchronized via readiness signals between jobs on HTCondor.  
For the server configurations with 4 and 40 GPUs, a stable throughput is observed, while for the 80 GPU configuration, a 1\% drop in throughput is observed.
This is the result of a small number of clients failing from gRPC errors that will be investigated in further studies.
The results of the sustained test showing throughput as a function of time are shown in Fig.~\ref{fig:hepcloud_throughput}.

\section{Data Availability}
Upon request, the corresponding author will provide data required to reproduce the figures.

\section{Code Availability}

The code from this work can be found at \url{http://github.com/fastmachinelearning/gw-iaas}.

%\bibliographystyleNew{aasjournal}
%\bibliographyNew{references}

\bibliographystyle{aasjournal}
\bibliography{references}

\end{document}